\documentclass[osajnl,twocolumn,showpacs]{revtex4}

\usepackage{graphicx}
\graphicspath{{pict/}{}}

\newcommand\pictc[5]{\begin{figure}
                   \centerline{
\includegraphics[width=#1\columnwidth,height=0.7\textheight,keepaspectratio]{#3}}
               \protect\caption{\protect\label{fig:#4} #5}
                \end{figure}            }

\newcommand\pict[4][1]{\pictc{#1}{!tb}{#2}{#3}{#4}}
\newcommand\rpict[1]{\ref{fig:#1}}

\newcommand\leqt[1]{\protect\label{eq:#1}}
\newcommand\reqtn[1]{\ref{eq:#1}}
\newcommand\reqt[1]{(\reqtn{#1})}

\newcounter{Fig}

\begin{document}
\begin{sloppy}

\title{Self-trapping of polychromatic light in nonlinear photonic lattices}

\author{Kristian Motzek}
\author{Andrey A. Sukhorukov}
\author{Yuri S. Kivshar}

\affiliation{Nonlinear Physics Centre and Centre for Ultra-high
bandwidth Devices for Optical Systems (CUDOS), Research School of
Physical Sciences and Engineering, Australian National University,
Canberra ACT 0200, Australia}

\begin{abstract}
We study dynamical reshaping of polychromatic beams due to
collective nonlinear self-action of multiple-frequency components in
periodic photonic lattices and predict the formation of {\em
polychromatic discrete solitons} facilitated by localization of
light in spectral gaps. We show that the self-trapping efficiency
and structure of emerging polychromatic gap solitons depends on the
spectrum of input beams due to the lattice-enhanced dispersion,
including the effect of crossover from localization to diffraction
in media with defocusing nonlinearity.
\end{abstract}


\maketitle

The fundamental physics of periodic photonic structures is governed
by the wave scattering from periodic modulations of the refractive
index and subsequent wave interference. Such a resonant process is
sensitive to a variation of the beam frequency and propagation
angle~\cite{Russell:1995-585:ConfinedElectrons}. Accordingly,
refraction and diffraction of optical beams may depend strongly on
the optical wavelength, allowing for construction of superprisms
that realize a spatial separation of the frequency components.

In this Letter we address an important question of how the
periodicity-enhanced sensitivity of diffraction upon wavelength
influences nonlinear self-action of polychromatic light. We show
that interaction between multiple-frequency components of an optical
beam can lead to a collective self-trapping effect and polychromatic
solitons, where spatial diffraction is suppressed simultaneously in
a broad spectral region. These solitons can exist in periodic
structures with noninstantaneous nonlinear response, such as
optically-induced lattices~\cite{Efremidis:2002-46602:PRE, Fleischer:2003-23902:PRL,
Neshev:2003-710:OL} or waveguide arrays~\cite{Chen:2005-4314:OE, Matuszewski:2006-254:OE} in photorefractive materials. We demonstrate that the spectrum of polychromatic solitons possesses a number of distinctive features, related to the structure of the photonic bandgap spectrum. This suggests the possibility to perform nonlinear probing and
characterization of the bandgap spectrum in the frequency domain,
extending the recently demonstrated approach for nonlinear
Bloch-wave spectroscopy with monochromatic
light~\cite{Bartal:2005-163902:PRL}.

We study the dynamics of polychromatic light in planar nonlinear
photonic structures with a modulation of the refractive index along
the transverse spatial dimension, such as optically-induced
lattices~\cite{Efremidis:2002-46602:PRE, Fleischer:2003-23902:PRL,
Neshev:2003-710:OL} or periodic waveguide
arrays~\cite{Chen:2005-4314:OE, Matuszewski:2006-254:OE}. Then, the evolution of polychromatic beams in media with slow 
nonlinearity can be described by a set of normalized nonlinear
equations,
\begin{equation} \leqt{nls}
   i \frac{\partial A_n}{\partial z}
   + \frac{\lambda_n z_0}{4 \pi n_0 x_0^2}
        \frac{\partial^2 A_n}{\partial x^2}
   + \frac{2 \pi z_0}{\lambda_n} \left[\nu(x)+\gamma I\right] A_n
   = 0\, ,
\end{equation}
where $A_n$ are the envelopes of the different frequency components
of vacuum wavelengths $\lambda_n$, $x$ and $z$ are the transverse
and longitudinal coordinates normalized to $x_0=10 \mu m$ and
$z_0=1mm$, respectively, $I=\sum_{n=1}^N |A_n|^2$ is the total
intensity, $N$ is the number of components,  $n_0$ is the average
refractive index, $\nu(x)$ is the refractive index
modulation in the transverse spatial dimension, and $\gamma$ is the
nonlinear coefficient. We consider the case of a Kerr-type medium
response, where the induced change of the refractive index is
proportional to the light intensity and neglect higher-order
nonlinear effects such as saturation, in order to clearly identify
the fundamental phenomena independent on particular nonlinearity. We
note that Eq.~\reqt{nls} with $\lambda_n = \lambda$ describe
one-color multigap solitons~\cite{Cohen:2003-113901:PRL,
Sukhorukov:2003-113902:PRL, Buljan:2004-223901:PRL,
Motzek:2005-2916:OE}.

\pict{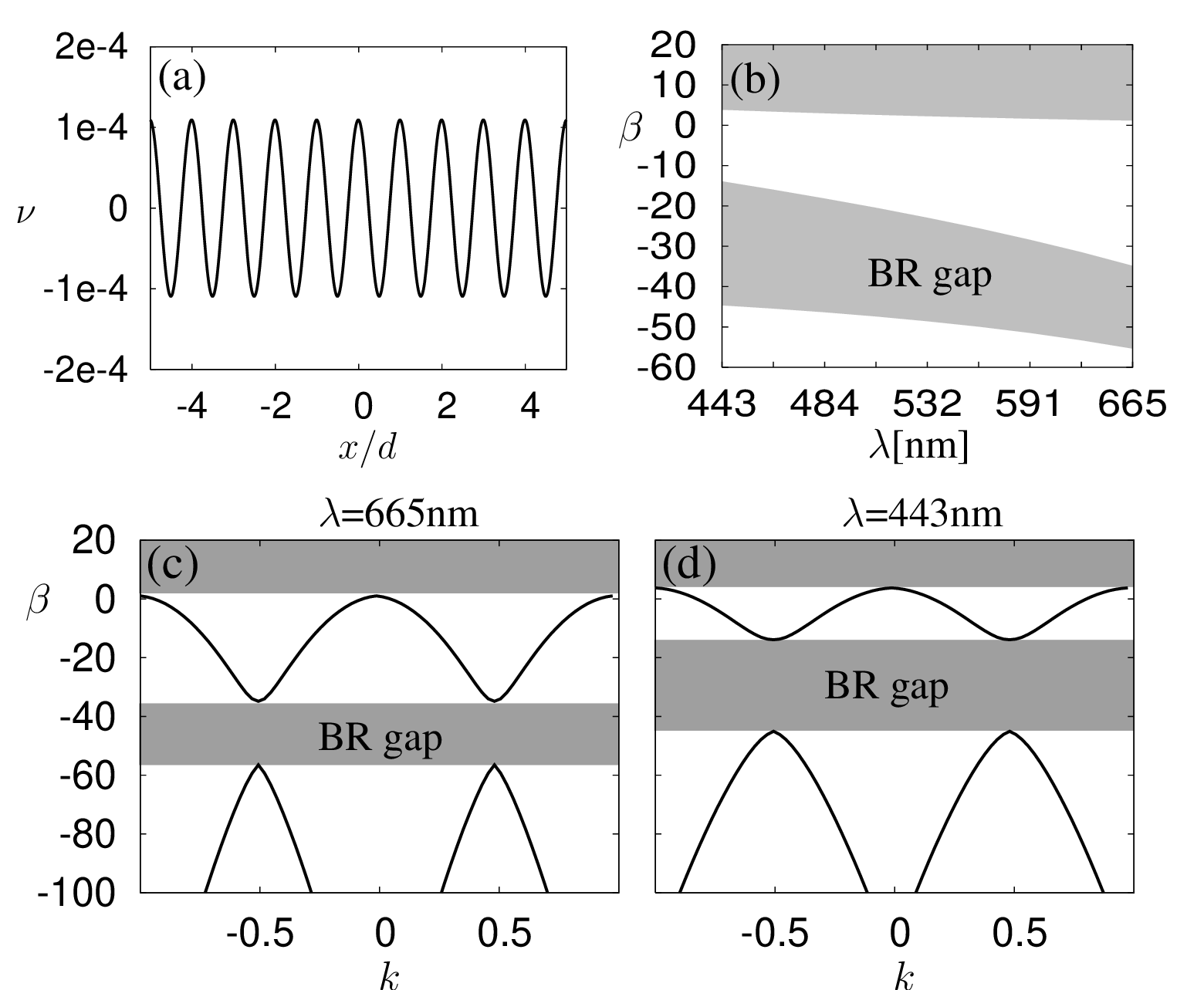}{bands}{ (a)~Refractive index contrast in a
lattice; (b)~Dependence of the bandgap spectrum on the wavelength,
and (c,d)~corresponding spatial Bloch-wave dispersion for two
different wavelengths, 665nm and 443nm, respectively. Transverse Bloch wavevector component $k$ is normalized to $K = 2 \pi / d$. Grey shading marks spectral gaps where waves become exponentially localized:
semi-infinite gap at the top (large $\beta$) and Bragg-reflection
(BR) gaps at smaller $\beta$.}

Linear dynamics of optical beams propagating in a periodic photonic
lattice is defined through the properties of extended eigenmodes
called Bloch waves~\cite{Russell:1995-585:ConfinedElectrons}. We
consider an example of lattice with $\cos^2$ refractive index
modulation [see Fig.~\rpict{bands}(a)] with the period $d=10\mu$m,
and calculate dependencies between the longitudinal ($\beta$, along $z$) and transverse ($k$, along $x$) wave-numbers for Bloch waves, see
Figs.~\rpict{bands}(b-d). The top spectral gap is semi-infinite
(extends to large $\beta$), and it appears due to the effect of the
total internal reflection. The effective diffraction of Bloch waves
becomes anomalous at the upper edges of Bragg-reflection gaps, where
$D_{\rm eff} = - \partial^2 \beta / \partial k^2 < 0$.

It is known that the presence of Bragg-reflection gaps and
associated anomalous diffraction regions allows for the formation of
monochromatic spatial gap solitons even in media with
self-defocusing nonlinearity~\cite{Kivshar:1993-1147:OL,
Fleischer:2003-23902:PRL, Matuszewski:2006-254:OE}. Results in Figs.~\rpict{bands}(b-d) show
that the spatial bandgap spectrum depends on the optical wavelength
and, in particular, we find that the {\em anomalous diffraction
regime is strongly frequency dependent} as $D_{\rm eff} \sim
\lambda^3$ at large wavelengths, whereas the bulk diffraction
coefficient is proportional to $\lambda$. Accordingly, the
Bragg-reflection gap becomes much narrower at larger wavelengths,
limiting the maximum degree of spatial localization that is
inversely proportional to the gap width.

The variation of the gap width can have a dramatic effect on
self-action of an input Gaussian beam focused at a single site of a
defocusing nonlinear lattice~\cite{Matuszewski:2006-254:OE}, where a
sharp crossover from self-trapping to defocusing occurs as the gap
becomes narrower. We note that, most remarkably, these distinct
phenomena can be observed in the same photonic structure but for
different wavelength components. In our numerical simulations, we
put $\gamma=-10^{-4}$ and choose the lattice parameters such that the critical wavelength
corresponding to the crossover is around 591nm. We confirm that the
monochromatic beam with $\lambda=443nm$ experiences strong
self-trapping, whereas the largest fraction of input beam power
becomes delocalized at a shorter wavelength $\lambda=665nm$. We then
address a key question of how an interplay between these opposite
effects changes the nonlinear propagation of polychromatic beams.

\pict{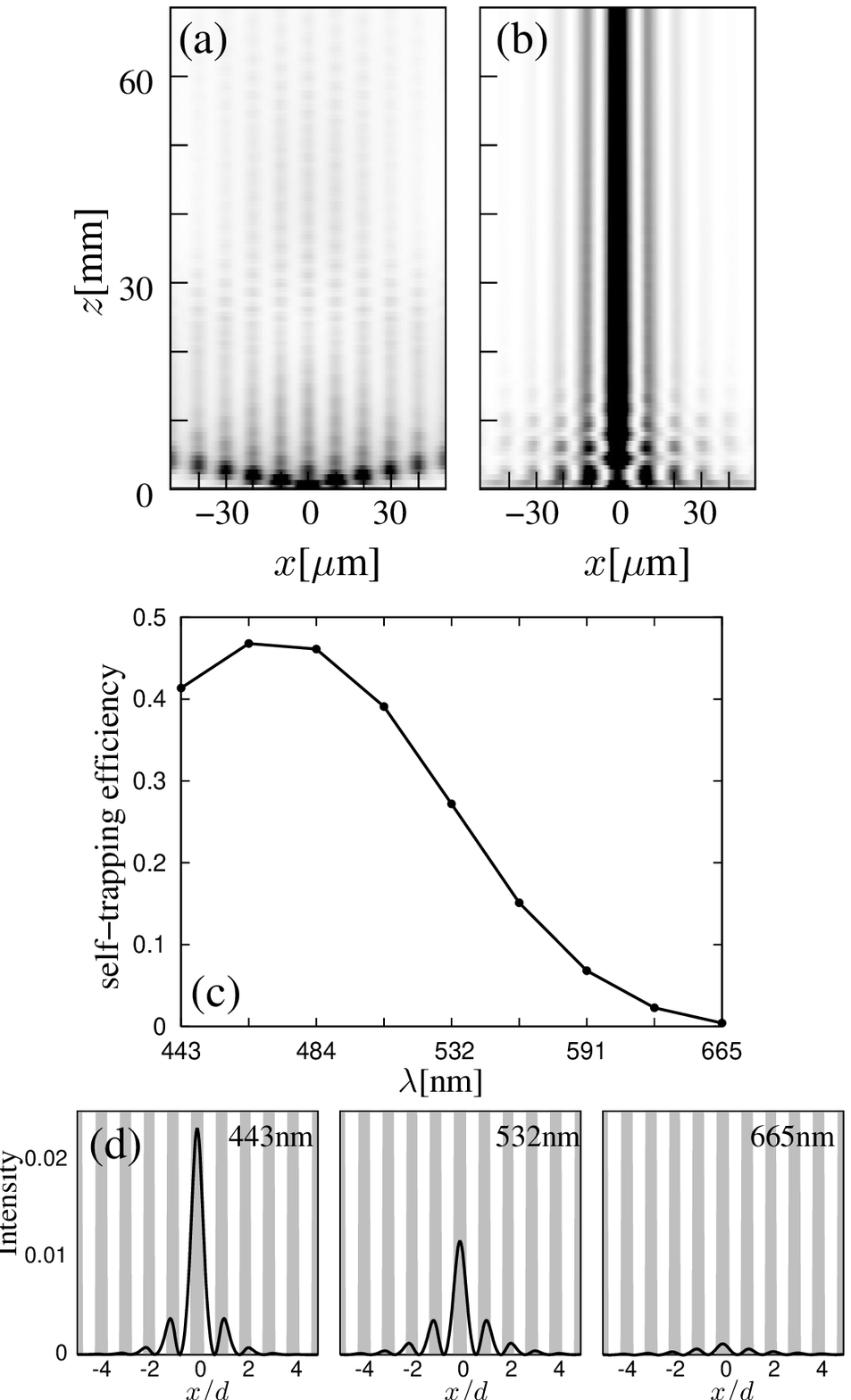}{generation}{ (a)~Linear diffraction of a
polychromatic beam; darker shading marks higher intensity;
(b-d)~Nonlinear self-focusing and the generation of a polychromatic
gap soliton for the peak input intensity $I = 1.11$: 
(b)~evolution of the total intensity, (c)~power
spectrum of the self-trapped soliton at the output, and (d)~output
intensity profiles of individual soliton components with wavelengths
$\lambda = 443$nm, 532nm, and 665nm.}

We model the self-action of polychromatic light beams by simulating
the propagation of nine components with the wavelengths ranging
between 443nm and 665nm. The input corresponds to a narrow Gaussian
beam that has the width of one lattice site, i.e. in our case
5$\mu$m. Figure~\rpict{generation} shows our numerical results for
the propagation of polychromatic light over 70mm. The spectrum of the
light at the input is \lq white\rq , i.e. the light beams of
different wavelength all have the same input profile and intensity.
In the linear regime (small input intensity), all components of the
beam strongly diffract, and the beam broadens significantly at the
output, as shown in Fig.~\rpict{generation}(a). As the input power
is increased, we find that the spatial spreading can be compensated
in a broad spectral region by self-defocusing nonlinearity. We
observe a spatially localized total intensity profile at the output,
indicating the formation of {\em a polychromatic gap soliton}
[Fig.~\rpict{generation}(b)].

We note that the spatial localization of the soliton components
strongly depends on the wavelength [Figs.~\rpict{generation}(d)], so
that the long wavelength component has a much larger spatial extent
than the short wavelength component. Hence, the soliton has a blue
center and red tails, and this effect is more pronounced than for
solitons with the same spectra in bulk media. Additionally, the
power spectrum of the soliton becomes blue-shifted at the output.
Figure~\rpict{generation}(c) shows the so-called \lq self-trapping
efficiency\rq, which we define here as the percentage of light that
remains in the three central waveguides of the optical lattice after
the propagation for each wavelength. This value essentially is
identical to the trapped fraction of light, as even for longer
propagation distances the light would remain localized in these
waveguides. We see that more that 40\% of the light with the
wavelengths between 443nm and 484nm is trapped, whereas for the
longer wavelengths that percentage drastically decreases due to the
narrowing of the Bragg-reflection gap. However, due to the nonlinear
interaction between the different wavelengths, there still is a
noticeable amount of light from the red side of the spectrum that is
trapped (roughly 8\% of the light at 591nm). This is in a sharp
contrast to the case of monochromatic red light propagation, where
the self-trapping efficiency vanishes.

\pict{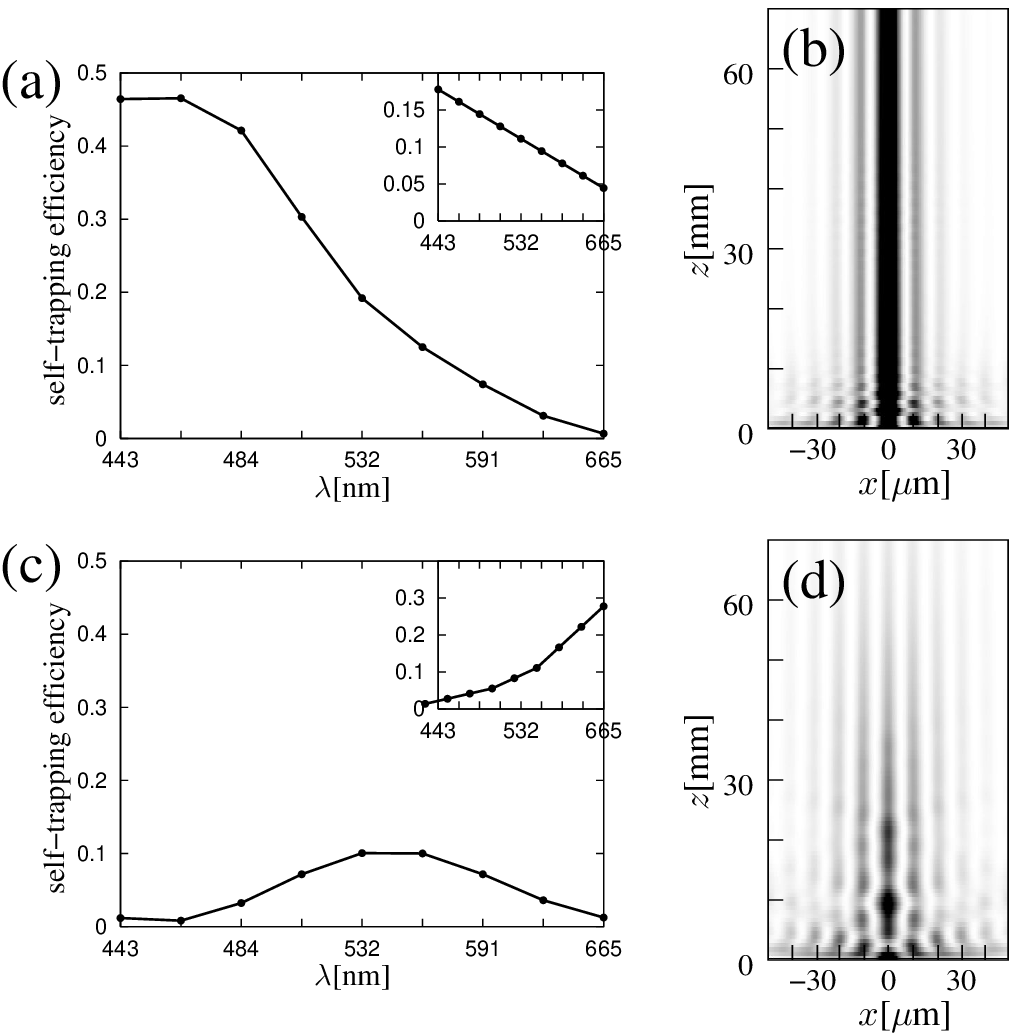}{spectrTransform}{ (a,c)~Fraction of the
light trapped by the lattice as a function of wavelength; (b,d)~the
corresponding evolution of the total intensity profiles for incident
light beams with the same input profiles, but different
frequency spectra shown in the insets.  The peak input intensity is (a,c)~the same or (b,d)~twice as big compared to Figs.~\rpict{generation}(b-d).}

We now study the effect of input frequency spectrum on the nonlinear
self-action of polychromatic light. We perform numerical simulations
for the same profiles of the input beam as in
Fig.~\rpict{generation}(b), but considering different power
distribution between the frequency components. 
Figures~\rpict{spectrTransform}(a-d) show the characteristic propagation results for beams with blue- and red-shifted input spectra.

For the blue-shifted input spectrum
[Figs.~\rpict{spectrTransform}(a,b)], we observe self-trapping of
the polychromatic light beam, and a small percentage of the red
light is trapped by the nonlinear index change caused by the blue
parts of the spectrum. In fact, the self-trapping efficiency for the
red part of the spectrum is almost identical to the case of a white
spectrum shown in Fig.~\rpict{generation}(c). Fundamentally
different behavior is observed for a polychromatic beam with
red-shifted spectrum [Figs.~\rpict{spectrTransform}(c,d)]. In this
case, the beam strongly diffracts and self-trapping does not occur even when the total input intensity is increased several times compared to the case of white spectrum. This happens due to the tendency of red components to experience enhanced diffraction as the effect of defocusing nonlinearity is increased at higher intensities.
We note that, according to Fig.~\rpict{spectrTransform}(c), the blue part of the spectrum is also diffracting. 

In conclusion, we have studied the propagation of polychromatic
light and the formation of polychromatic solitons in periodic
photonic lattices, and demonstrated that light self-action can be
used to reshape multiple frequency components of propagating beams
in media with noninstantaneous nonlinear response, such as
photorefractive materials or liquid crystals. We have demonstrated
that self-trapping efficiency and structure of emerging
polychromatic gap solitons depends strongly on the spectrum of input
beams due to the lattice-enhanced dispersion, and identified the
effect of crossover between localization and diffraction in
defocusing media.

\end{sloppy}

\begin{thebibliography}{10}

\bibitem{Russell:1995-585:ConfinedElectrons}
P.~St.~J. Russell, T.~A. Birks, and F.~D. Lloyd~Lucas, ``Photonic Bloch waves
  and photonic band gaps,''  in {\em Confined Electrons and Photons}, E.
  Burstein and C. Weisbuch, eds., (Plenum, New York, 1995), \ pp.\ 585--633.

\bibitem{Efremidis:2002-46602:PRE}
N.~K. Efremidis, S. Sears, D.~N. Christodoulides, J.~W. Fleischer, and M.
  Segev, ``Discrete solitons in photorefractive optically induced photonic
  lattices,'' Phys. Rev. E {\bf 66,} 046602--5 (2002).

\bibitem{Fleischer:2003-23902:PRL}
J.~W. Fleischer, T. Carmon, M. Segev, N.~K. Efremidis, and D.~N.
  Christodoulides, ``Observation of discrete solitons in optically induced real
  time waveguide arrays,'' Phys. Rev. Lett. {\bf 90,} 023902--4 (2003).

\bibitem{Neshev:2003-710:OL}
D. Neshev, E. Ostrovskaya, Y. Kivshar, and W. Krolikowski, ``Spatial solitons
  in optically induced gratings,'' Opt. Lett. {\bf 28,} 710--712 (2003).

\bibitem{Chen:2005-4314:OE}
F. Chen, M. Stepic, C.~E. Ruter, D. Runde, D. Kip, V. Shandarov, O. Manela, and
  M. Segev, ``Discrete diffraction and spatial gap solitons in photovoltaic
  {{LiNbO}$_3$} waveguide arrays,'' Opt. Express {\bf 13,} 4314--4324 (2005).

\bibitem{Matuszewski:2006-254:OE}
M. Matuszewski, C.~R. Rosberg, D.~N. Neshev, A.~A. Sukhorukov, A. Mitchell, M.
  Trippenbach, M.~W. Austin, W. Krolikowski, and Yu.~S. Kivshar, ``Crossover
  from self-defocusing to discrete trapping in nonlinear waveguide arrays,''
  Opt. Express {\bf 14,} 254--259 (2006).

\bibitem{Bartal:2005-163902:PRL}
G. Bartal, O. Cohen, H. Buljan, J.~W. Fleischer, O. Manela, and M. Segev,
  ``Brillouin zone spectroscopy of nonlinear photonic lattices,'' Phys. Rev.
  Lett. {\bf 94,} 163902--4 (2005).

\bibitem{Cohen:2003-113901:PRL}
O. Cohen, T. Schwartz, J.~W. Fleischer, M. Segev, and D.~N. Christodoulides,
  ``Multiband vector lattice solitons,'' Phys. Rev. Lett. {\bf 91,} 113901--4
  (2003).

\bibitem{Sukhorukov:2003-113902:PRL}
A.~A. Sukhorukov and Yu.~S. Kivshar, ``Multigap discrete vector solitons,''
  Phys. Rev. Lett. {\bf 91,} 113902--4 (2003).

\bibitem{Buljan:2004-223901:PRL}
H. Buljan, O. Cohen, J.~W. Fleischer, T. Schwartz, M. Segev, Z.~H. Musslimani,
  N.~K. Efremidis, and D.~N. Christodoulides, ``Random-phase solitons in
  nonlinear periodic lattices,'' Phys. Rev. Lett. {\bf 92,} 223901--4 (2004).

\bibitem{Motzek:2005-2916:OE}
K. Motzek, A.~A. Sukhorukov, F. Kaiser, and Yu.~S. Kivshar, ``Incoherent
  multi-gap optical solitons in nonlinear photonic lattices,'' Opt. Express
  {\bf 13,} 2916--2923 (2005).

\bibitem{Kivshar:1993-1147:OL}
Yu.~S. Kivshar, ``Self-localization in arrays of defocusing wave-guides,'' Opt.
  Lett. {\bf 18,} 1147--1149 (1993).

\end{thebibliography}
\end{document}